\documentclass[11pt,preprint,graphicx]{aastex}

\def\etal{\it et al. \rm }
\begin{document}

\title{Age and Metallicities of Cluster Galaxies: A1185 and Coma}

\author{Karl Rakos}
\affil{Institute for Astronomy, University of Vienna, A-1180, Wien, Austria;
karl.rakos@chello.at}

\author{James Schombert}
\affil{Department of Physics, University of Oregon, Eugene, OR 97403;
js@abyss.uoregon.edu}

\author{Andrew Odell}
\affil{Department of Physics and Astronomy, Northern Arizona University, Box
6010,
Flagstaff, AZ 86011; andy.odell@nau.edu}

\begin{abstract}

We present age and metallicities determinations based on narrow band
continuum colors for the galaxies in the rich clusters A1185 and Coma.
Using a new technique to extract luminosity-weighted age and [Fe/H] values
for non-star-forming galaxies, we find that both clusters have two separate
populations based on these parameters.  One population is old ($\tau >$ 11
Gyrs) with a distinct mass-metallicity relation.  The
second population is slightly younger ($\tau \approx$ 9 Gyrs) with lower
metallicities and lower stellar masses.  We find detectable correlations
between age and galaxy mass in both populations such that older galaxies
are more massive and have higher mean metallicities, confirming previous
work with line indices for the same type of galaxies in other clusters
(Kelson \etal 2006, Thomas \etal 2005).  Given the previously discovered
correlation between galaxy mass and $\alpha$/Fe abundance (a measure of the
duration of initial star formation, Denicolo \etal 2005, Sanchez-Blazquez
\etal 2006), we interpret our age-metallicity correlations to imply that
cluster galaxies are coeval with varying durations for their initial
bursts.   Our results imply shorter durations for higher mass galaxies, in
contradiction to the predictions of classic galactic wind models.  Since we
also find a clear mass-metallicity relation for these galaxies, then we
conclude that star formation was more efficient for higher mass galaxies, a
scenario described under the inverse wind models (Matteucci 1994).  With
respect to cluster environmental effects, we find there is a significant
correlation between galaxy mean age and distance from the cluster center,
such that older galaxies inhabit the core.  This relationship would
nominally support hierarchical scenarios of galaxy formation (younger age
in lower density regions); however, environmental effects probably have
larger signature in the sample and present-day galaxies are remnants from
an epoch of quenching of initial star formation, which would result in the
same age gradients.

\end{abstract}

\keywords{galaxies: evolution --- galaxies: stellar content ---
galaxies: elliptical}

\section{INTRODUCTION}

The key to understanding the star formation history of galaxies is
determining the age of their underlying stellar populations and the amount
chemical evolution that those populations have undergone.  Historically,
these issues have been addressed separately and by morphological type.  For
example, early-type galaxies are considered have the simplest modes of star
formation, the oldest ages (i.e. primordial) and relatively passive
evolution as indicated by their similarity in color, morphology and
kinematic properties (Bernardi \etal 2003, Tremonti \etal 2004, Cool \etal
2006, Smolcic \etal 2006).  Very little current or recent star formation
(Bernardi \etal 2001), plus low amounts of molecular gas (Huchtmeier, Sage
\& Henkel 1995), reinforces the hypothesis that a majority of their stars
have their origin at high redshift (Larson
1975, Kodama \etal 1998, Jimenez \etal 1999, van Dokkum \& Stanford 2003).

Work on the star formation history of galaxies is dominated by studies of
rich clusters because 1) they are located in the same region of position
and velocity space easing observational constraints, 2) they are easily
detected in surveys at large redshift and are useful signposts for the
study of evolutionary effects and 3) they are impacted by similar, if not
complicated, environmental effects.  Since rich clusters are dominated by
elliptical and S0 galaxies, then they are ideal laboratories for the
understanding of the simplest systems, but the discovery of the
Butcher-Oemler effect (Butcher \& Oemler 1984, Poggianti 2004) means that
rich clusters are also interesting arenas to study starburst
phenomenon.

Tracers of age and metallicity in even simple systems, such as cluster
ellipticals, are complicated by the fact that, unlike globular clusters,
the stellar populations in ellipticals, are a mixture of stars of many ages
at various stages of chemical evolution (see studies that resolve the
stellar populations in ellipticals by HST, Gregg \etal 2004).  This means
that any information obtained on galaxies, outside the resolution range of
space imaging, will be luminosity weighted and observational constraints
are such that these data must be interpreted using the guidance of star
formation models.  While model guided results always run the gamut of
dependences on various assumptions, the range of galaxy types and lookback
time has successful allowed a critical examination of these assumptions
(Renzini 2006).  In addition, the star formation and chemical evolution
histories of galaxies can be examined through the use of many parameters
based on their integrated light (e.g.  broadband colors versus line
spectroscopy), each providing their own unique picture of the underlying
stellar populations.

The determination of a galaxy's age and metallicity (meaning not only the
mean quantities, but the distribution of age's and metallicities within a
galaxy's stellar population) is an observationally intense problem and
suffered in early studies from the problem that broadband colors are
coupled with respect to age and metallicity (the so-called
age-metallicity degeneracy problem, see Worthey 1999).  However, in the
last decade, three approaches have successful broken the age-metallicity
degeneracy.  The first is the use of colors across a wide wavelength region
(e.g.  far-UV to near-IR, see Kaviraj \etal 2006) which bridges regions of
the spectrum sensitive to many different regions of the HR diagram.
The second method is to use various spectral features sensitive to
age and metallicity in a semi-independent manner (Gonzalez 1993).
Semi-independent in a fashion that meaningful values can only be extracted by
comparison to stellar population models.  There is nothing new to the use
of models to determine stellar population parameters, for example, globular
cluster ages can only be extracted by comparison to stellar evolution
tracks.  But an additional complications arises in that the stellar
populations in galaxies are neither homogeneous in age nor metallicity
(if our own Galaxy is a typical example).  The third technique is
the one pioneered by our team using narrow band filters combined
with a robust principal component analysis (PCA) of multi-color space
derived from spectroevolutionary (SED) models (Rakos \& Schombert 2005a).  This
technique makes use of the effects of age and metallicity on the continuum
emission resulting from the integrated light of the underlying stellar
population.

For the past decade, the technique of choice to determine age and
metallicity in galaxies has been judicious choice of certain spectral line
indices (i.e. the Lick sequence, Worthey \& Ottaviani 1997, Trager \etal
2000).  These studies sample particular lines sensitive to metallicity
(e.g., Fe and Mg) combined with other couplets to find age (e.g. the Balmer
lines).  The grid of age and [Fe/H] is model dependent, but the two values
are distinct and have a solvable degeneracy problem.  While line
measurements would appear, at first glance, to be a superior technique
since their are actually measuring the parameters to investigate the
chemical evolution of stellar populations (e.g. Fe, Mg), line strengths do
suffer a number of disadvantages.  Increased information on various lines
paints a confusing picture.  For example, Fe strengths show very little
dependence on galaxy mass (Fisher, Franx \& Illingworth 1997), in contrast
to expectations from the color-magnitude relation.  There are sharp
dependences between $\alpha$ elements, such as Ca and Mg, and the main
source of free electrons in stellar photospheres, Fe.  For example,
Sanchez-Blazquez \etal (2006b) find no correlation between age and
metallicity when the Mg$_2$ index is used, but clear correlations are
found with Ca or Fe indices.  This leads to difficulties in forming an
overall picture of galaxy formation as any model of stellar populations
does not necessarily desire the value of individual elements as input
parameters, but rather their combined effect on stellar atmospheres as this
reflects into their position on the HR diagram.

Recently, the values for galaxy age extracted from the Lick technique has
come in conflict with age determination by other methods.  Numerous studies
of cluster ellipticals using the Lick system that find many examples of
ellipticals with relatively young mean ages (less than 6 Gyrs,
Sanchez-Blazquez \etal 2006).  In contrast, our previous work using
continuum colors, flux values determined from regions of the near-blue
portion of a galaxy's spectra, find very few ellipticals with ages less
than 10 Gyrs.  And recently, ellipticals have also been determined to have
old ages in a study using far-IR indicators of age (Bregman, Temi \&
Bregman 2006).  Some of age dispersion may be due to contributions from
small amounts of young stars, significant in their effect on colors, but
minor in their fraction compared to the total stellar population (i.e., the
`frosting' effect, Trager \etal 2000).  However, again, this will be an
advantage for integrated continuum colors which more fully take into
account the whole stellar population in a galaxy and will not suffer from
aperture effects that may distort spectral estimates.

The dispersion in ages for cluster galaxies is a major inhibitor in
determining the correct scenario for galaxy formation and evolution.  While
a range of ages can be supported by various methods of galaxy formation,
purely old galaxy ages imply high redshifts for galaxy formation and
relatively simple evolutionary scenarios.  To further explore the issue of
galaxy age, we developed a technique parallel to line indices work that
focuses on the continuum region of the spectra around 4000\AA.  We have
refined our calculations of age and metallicity from our narrow band colors
and incorporated a robust technique to determine those values from the data
(see Rakos \& Schombert 2005a).  This paper will apply these new methods to
examine the history of galaxies in a nearby rich cluster, A1185, as
compared to our typical rich cluster for extragalactic studies, Coma.  Our
goal is to further refine the spectrophotometric classification of galaxies
on our system and determine the age and metallicity properties of cluster
galaxies with respect to their stellar mass and cluster environment.

\section{OBSERVATIONS}

The rich cluster A1185 is a Bautz-Morgan type II cluster with a richness
class of 1.  It is located at a redshift of 0.033 and is approximately 80
arcmins in diameter.  A1185 is a dynamically young cluster based on its
projected density and velocity distribution.  It composed of several distinct
(spatially as well as kinematically) subgroups and a core which has
several subclumps (Mahdavi \etal 1996).  Our photometry includes only those
galaxies on the region that is one Mpc in radius centered at the x-ray
luminosity peak.  This region also contains the top 10 brightest cluster
members and will, presumable, become the cluster core when virialization is
complete.  A1185 was observed on 08 Jan 2005, using the 2.3m Bok telescope
of the Steward Observatory located at Kitt Peak Arizona.  The image
device was 90prime, a prime focus wide-field imager using a mosaic of four
4k by 4k CCDs resulting in a field of view of 1.16 by 1.16 degrees and a
plate scale of 0.45"/pixel (Williams \etal 2004).

The filter system, a modified Str\"omgren ($uvby$) system, was the same as
used in our distant cluster studies (Rakos \& Schombert 1995).  Five
exposures of 600 secs were made in each filter centered on NGC 3552.
Calibration was obtained through a number of spectrophotometric standards
measured on each night.  The modified Str\"omgren system (herein called
$uz,vz,bz,yz$) is altered in the sense that the filters are slightly
narrower (by 20\AA) and the $u$ filter is shifted 30\AA\ to the red in its
central wavelength as compared to the original system.  The system we use
herein is called the $uz,vz,bz,yz$ system to differentiate it from the
original $uvby$ system since our filters are specific to the rest frame of
the cluster that is being studied.  The $uz,vz,bz,yz$ system covers three
regions in the near-UV and blue portion of the spectrum that make it a
powerful tool for the investigation of stellar populations in SSP's (simple
stellar population), such as star clusters, or composite systems, such as
galaxies.  The first region is longward of 4600\AA, where in the influence
of absorption lines is small.  This is characteristic of the $bz$ and $yz$
filters ($\lambda_{eff}$ = 4675\AA\ and 5500\AA), which produce a
temperature color index, $bz-yz$.  The second region is a band shortward of
4600\AA, but above the Balmer discontinuity. This region is strongly
influenced by metal absorption lines (i.e. Fe, CN) particularly for
spectral classes F to M, which dominate the contribution of light in old
stellar populations.  This region is exploited by the $vz$ filter
($\lambda_{eff} = 4100$\AA).  The third region is a band shortward of the
Balmer discontinuity or below the effective limit of crowding of the Balmer
absorption lines.  This region is explored by the $uz$ filter
($\lambda_{eff} = 3500$\AA).  All the filters are sufficiently narrow (FWHM
= 200\AA) to sample regions of the spectrum unique to the various physical
processes of star formation and metallicity (see Rakos \etal 2001 for a
fuller description of the color system and its behavior for varying
populations).

The reduction procedures have been published in Rakos, Maindl \& Schombert
(1996) and references therein. The photometric system is based on the
theoretical transmission curves of filters (which can be obtained from the
authors) and the spectra of spectrophotometric standard stars published in
the literature (Massey \& Gronwall, 1990).  The convolution of the
transmission curves and the spectra of the standard stars produces
theoretical flux values for color indices of the standard stars corrected
for all light losses in the equipment and the specific sensitivity of the
CCD camera.  Magnitudes were measured on the co-added images using standard
IRAF procedures and are based for brighter objects on metric apertures set
at 32 kpc for cosmological parameters of $H_o=75$ km s$^{-1}$ Mpc$^{-1}$
and the Benchmark cosmology ($\Omega_m=0.3$, $\Omega_\Lambda=0.7$).  For
fainter objects, the apertures were adapted to deliver the best possible
signal to noise ratio, but always using the same aperture for all four
filters.  Color indices are formed from the magnitudes: $uz-vz$, $bz-yz$,
$vz-yz$, and $mz$ [$=(vz-bz)-(bz-yz)$].  There is a small, but measurable
galactic extinction in the region around A1185 ($E(B-V)=0.029$ mag), which
corresponds to $E(uz-vz)=0.014$, $E(bz-yz)=0.021$ and $E(vz-yz)=0.034$.
While these corrections were made to all the datasets, their effect on the
age and metallicity results (below) were negligible and are dominated by
other sources of error.  Typical errors were 0.03 mag in $vz-yz$ at the
bright end of the sample and 0.08 mag at the faint end.

After applying our photometric membership criteria to remove foreground and
background galaxies (see Rakos \etal 1991 for a full description of the
procedure) there remained 192 galaxies ranging from the brightest at
$m_{5500}=13.51$ (NGC 3550) to a completeness limit of 20 mag.  Of those
192, 53 had known redshifts and all 53 galaxies were cluster members
(meaning their redshifts were within 1000 km/sec of the cluster mean
redshift).  This confirms the high reliability of our photometric
membership technique, as discussed in our previous papers.  Six galaxies in
our sample display radio emission and one galaxy is a LINER.  Our
photometry system uses filters that avoid strong emission lines of star
formation or LINER activity, plus our colors are based on integrated colors
where the stellar population dominates over any nuclear emission.  This is
demonstrated by our colors for the above mentioned galaxies which are
normal elliptical types under our photometric classification methods.

\section{DISCUSSION}

\subsection{Colors and Photometric Morphology}

The three colors ($uz-yz$, $vz-yz$ and $bz-yz$) and the $mz$ index ($mz =
(vz-bz) - (bz-yz)$) for all 192 galaxies in A1185 are displayed in Figure
1.  This diagram is the main diagnostic tool for interpreting the narrow
band colors.  It has the advantage of demonstrating that the continuum
colors of galaxies is a relatively simple, to first order, distribution
over a range of Hubble types.  The distribution of colors in Figure 1 is
extremely well correlated, although the typical error is less than the
width of the correlation.  Changes in metallicity are a linear relation in
color space (for a constant age, see solid line for a 13 Gyr population),
thus, the scatter about a mean relationship suggests the effects of age for
the red population (see below) and recent star formation for the blue
population.  As noted by Smolcic \etal (2006), the narrow band colors of
galaxies are nearly a one parameter family and that the locus width in
principal component space is well explained by simple star formation
models.  These colors are also strongly correlated with spectral indices,
even emission lines, which track the current and recent star formation
histories.

\begin{figure}
\centering
\includegraphics[width=17cm]{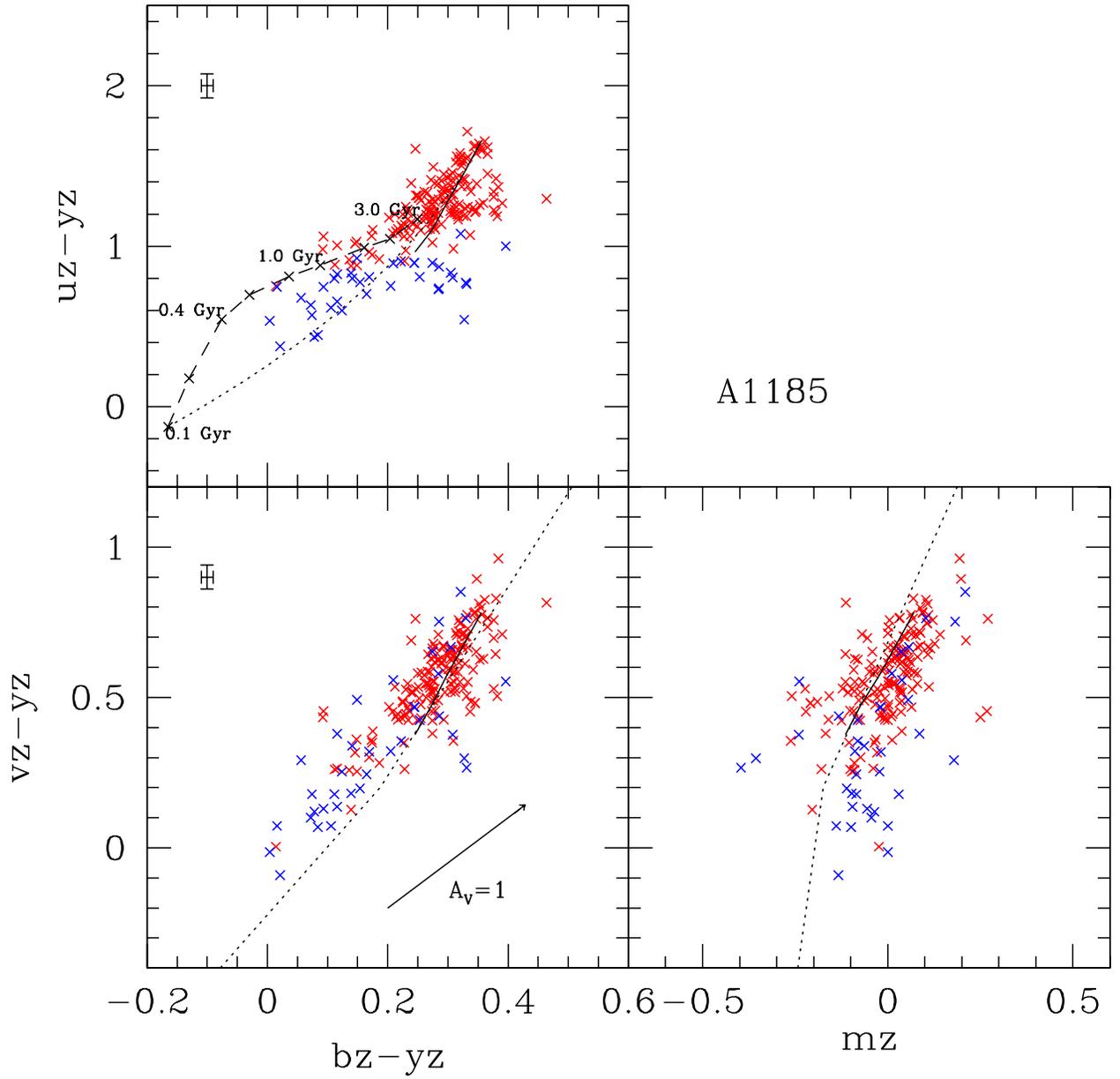}
\caption{Multi-color diagrams for A1185.   All three colors and the $mz$
index ($mz = (vz-bz) - (bz-yz)$) are shown for galaxies determined to be
cluster members by photometric criteria.  
The red population (galaxies with $bz-yz > 0.22$) is denoted by that
color as is the blue population.  
The solid line in each diagram displays the 13 Gyr models
for metallicities ranging from $-$0.7 to +0.4 (Schulz \etal 2002).  The dotted line in the
$vz-yz$ and $mz$ diagrams represents the 99,000 galaxies from the SDSS
sample Smolcic et al. (2006).  The dashed line in the $uz-yz$ diagram is
the range in age for a solar metallicity model.  The dotted line in the
same diagram displays a `frosting' model, i.e. adding increasing fractions
of a 0.1 Gyr population to a 13 Gyr population.  The reddening vector is
shown in the $vz-yz$ diagram.  The mean error bar for the sample is shown in the
upper left of each panel.}
\end{figure}

The uniform locus in colors is particularly important for our study as the
sample is selected from galaxies in a rich cluster.  The combination of
dense environment and a limited range in morphological types produces an
extremely red population of galaxies with a limited scatter.  Figure 1
displays the obvious result that a majority of the galaxies in A1185 have
typical elliptical colors ($bz-yz > 0.25$ and $uz-yz > 1.2$) and follow an
old (greater than 10 Gyrs) track of metallicity from spectroevolutionary
models (a 13 Gyr track is shown as a solid line in Figure 1 for
comparison).  We can also compare these colors with synthetic colors
generated from SDSS spectra (Smolcic \etal 2006).  Their mean color locus
are shown in Figure 1 as a dotted line (only for $vz-yz$ and $bz-yz$ as
SDSS spectra do not cover the $uz$ filter).  There is a small shift due to
slight different definitions of the filter bandpasses (also seen on other
comparison data, Rakos \& Schombert 2005b), but the slopes and dispersions
are identical.

A strongly defined color locus opens the possibility of classifying
galaxies by their spectrophotometric properties, in contrast to their
morphological appearance.  While classification by morphological appearance
has been a useful diagnostic for galaxy evolution, the underlying physics
is difficult to unravel.  Classification by optical color reflects, mostly,
the physics of the star formation histories of galaxies, which is
potentially more insightful to the evolution and formation of the stellar
populations in galaxies.

We have applied PCA in our previous papers (see Odell, Schombert \& Rakos
2002) to define four classes of galaxies based on their narrow band colors.
The most recent form of our classification system divides the first
principal component axis (PC1) into four subdivisions based on mean past
star formation rate; E (passive, red objects), S (star formation rates
equivalent to a disk galaxy), S- (transition between E and S) and S+
(starburst objects).  By comparison to SED models and nearby galaxies
(Rakos, Maindl \& Schombert 1996), the divisions along the PC1 axis are
drawn such that S galaxies correspond to those systems with spiral star
formation rates (approximately 1 $M_{\sun}$ per yr) and S+ galaxies
correspond to starburst rates (approximately 10 $M_{\sun}$ per yr).  We
note that since these divisions are determined by continuum colors, versus
spectral lines, they do not represent the current star formation rate, but
rather the mean star formation rate averaged of the last few Gyrs as
reflected into the optical emission by the dominant stellar population.
The red E systems display colors with no evidence of star formation in the
last five Gyrs.  The transition objects, S-, represents the fact that there
is not sharp division between the E class and S class .  These objects
display slightly bluer colors (statistically) from the passive E class;
however, the difference could be due to a recent, low-level burst of star
formation or a later epoch of galaxy formation or an extended phase of
early star formation or even lower mean metallicity (i.e.  the
color-magnitude effect).  In addition to classification by color, we can
separate out objects with signatures of non-thermal continuum (AGN) under
the categories of A+, A and A- based on their PC2 values colors.  It is
important to remember that these classifications are based solely on the
principal components as given by the color indices from four filters.
While, in general, these spectrophotometric classes map into morphological
ones (i.e.  E types are often ellipticals, S- are S0 and early-type
spirals, S are late-type spirals and S+ are irregulars), this system
differs from morphology by being independent of the appearance of the
galaxy and based on the color of the dominant stellar population in a
galaxy.  This is also a classification based on integrated colors, such
that large B/D galaxies (i.e. early type spirals) will generally be found
in class S- as the bulge light dominates over the disk.  This is in
contrast to classification by morphology where the existence of even a
faint disk distinguishes the galaxy from an elliptical.

Galaxies are labeled by their various spectrophotometric classes in Figure
1, where the size of the symbol is proportional to their absolute
luminosity.  Unsurprisingly, a majority of the galaxies are E class (48\%)
and S- (33\%).  The remaining 19\% were S class (15\%) and S+ (4\%).  The
Coma cluster (Odell, Schombert \& Rakos 2002) has a similar mix of S/S+ galaxies at 12
and 5\% each.  However, in Coma, a majority (72\%) of the red galaxies are
E class and 14\% are S- class, compared to 48\% E class in A1185 versus
33\% S- class.  It is important to remember that the S- class galaxies
represents a transition set of colors between star-forming (S class) and
non-star-forming (E class) as mapped into PC space.  While S- class
galaxies are, in general, associated with systems with low amounts of star
formation (large bulge disks, S0's and E+A), they are also prominent as one
examines lower luminosity systems where low mean metallicity pushes all the
narrow band colors to the blue (see Figure 3 below).

\begin{figure}
\centering
\includegraphics[width=17cm]{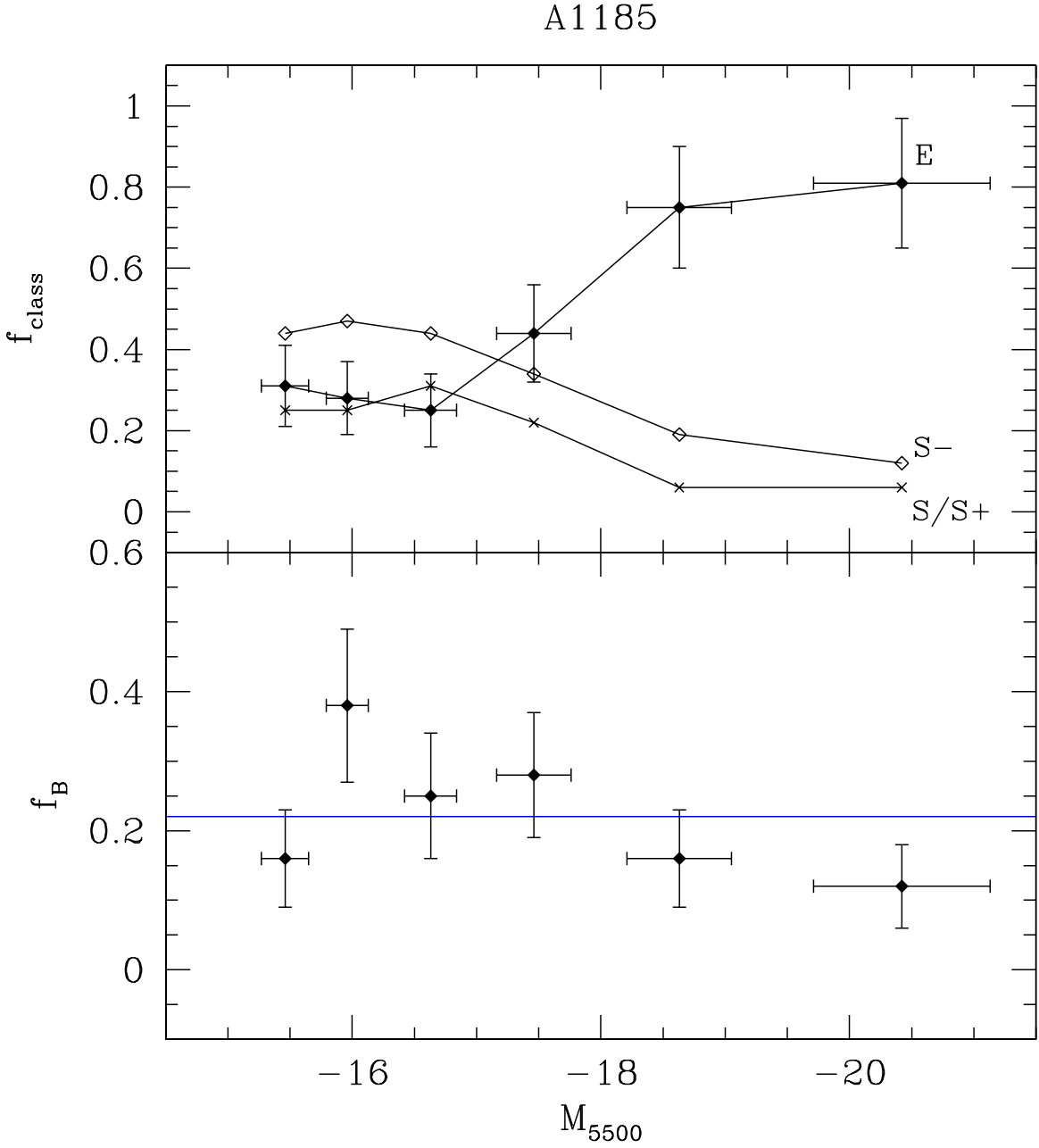}
\caption{The fraction of spectrophotometric types and the blue fraction (by
$bz-yz$ color) as a function of absolute magnitude ($M_{5500}$).  The drop in
the fraction of E class galaxies as a function of luminosity reflects the
color-magnitude relation for ellipticals rather any morphological effect.  The
increase in star forming galaxies (type S and S+) is real in that the blue
population in a rich cluster increases with decreasing luminosity.
}
\end{figure}

The higher fraction of S- class galaxies in A1185 (compared to Coma) is
consistent with the higher fraction of blue galaxies found in A1185.  The
total blue fraction ($f_B$) for A1185, defined as the ratio of the number
of galaxies with $bz-yz < 0.22$ (Rakos \etal 2000) is 0.21 (trimmed to the
same limiting absolute magnitude as Coma).  For comparison, the value for
Coma is 0.10.  Given that A1185 is a dynamically younger cluster than Coma,
a large blue fraction is to be expected (see review of cluster
environmental effects in Poggianti 2004).  A simple blue fraction value
disguises a great deal of information about the population of galaxies in a
cluster.  The bottom panel in Figure 2 displays the run of $f_B$ as a
function of absolute magnitude (divided into bins of 32 galaxies).  As with
other rich clusters we have studied (A2218, A2125 Rakos \& Schombert 2005b,
A115, A2283 Rakos \etal 2000, A2317 Rakos \etal 1997), a common trend of
increasing $f_B$ with fainter magnitudes is found.  While galaxies do
become bluer with decreasing luminosity (i.e., CMR), this effect is minimal
in the $bz-yz$ color where $f_B$ is defined (see below).  The top panel in
Figure 2 also shows that fraction of star-forming galaxies (S and S+ class)
also increases with decreasing luminosity and is the driving factor in
$f_B$.

One of the earliest photometric correlations for galaxies is the
color-magnitude relation (CMR) and A1185 is unique by having the deepest
examination of its CMR of any cluster of galaxies (Andreon \etal 2006).
The Andreon \etal study (using $BVR$ colors) found the CMR for A1185 to be
linear and well defined down to absolute luminosities of $M_V = -14$.
Figure 3 displays the CMR for all three narrow band colors in our study.
The solid lines are fits to the Coma data and match perfectly to the A1185,
even at the level of the scatter around the relationships.  The
universality of the CMR is well documented (see Schawinski \etal 2006 and
references therein), and its origin as a mass-metallicity relationship is
also well-studied by theoretical models (Kodama \& Arimoto 1997, although
see Trager \etal 2000 for a dissenting view).  While determining exactly
how the age and metallicity of a galaxy map into the CMR requires guidance
from SED models, it is known from our previous work in the Coma cluster
(Odell Schombert \& Rakos 2002) that the primary driver of galaxy colors is
metallicity, with age taking a secondary role if there is no recent star
formation.  The competition between age and metallicity is the main goal of
this paper and we take the existence of the mass-metallicity relation for
ellipticals to be given (Gallazzi \etal 2006).

With respect to our narrow band colors, there are series of additional
points.  First, the CMR is difficult to analyze as a linear relation due to
the fact that any star-forming galaxies lie on the blue side of the
relation which unbalances a linear fit.  Our spectrophotometric
classification system has the immediate advantage of removing star-forming
galaxies from the fit, however, one then finds increasing weight given to
red objects which biases the fit in the other direction.  In addition, the
separation between E and S- classes is arbitrary (although guided by
morphological information), thus, final fits will be dependent on the cut
used to separate the photometric categories.  Fortunately, in a rich
cluster, the number of red galaxies vastly outnumbers any contaminating
blue galaxies.  Experimentation with different cuts and bi-weight fitting
procedures finds that the results for A1185 are statistically identical to
CMR fits to other rich clusters in our previous papers.  The scatter on the
final slope and zeropoint shown in Figure 3 was due solely to observational
uncertainty in the colors rather than sample criteria.

A second difference between the Coma sample and A1185 is that the A1185
data involves a fainter magnitude cutoff.  We find that this resolves an
anomaly in the Coma $bz-yz$ CMR noted in Odell, Schombert \& Rakos (2002).
For the Coma data, we were surprised to find that the $bz-yz$ CMR had no
slope in a formal fit.  Our expectation from SED models was that a slope,
due to metallicity effects, in $uz-yz$ and $vz-yz$ should translate into a
smaller, but measurable slope in $bz-yz$.  In $vz-yz$, the CMR drops from a
mean of 0.80 for $M_{5500}=-22$ to approximately 0.50 at $-16$.  For a 13
Gyr population, the same change in $vz-yz$ (0.30) corresponds to a change
in 0.08 in $bz-yz$, which was not seen in the Coma data where a formal fit was
equivalent to a zero slope.

Lacking a clear CMR in $bz-yz$, we interpreted the lack of slope to be an
age effect countering the mass-metallicity effect (where $bz-yz$ is more
sensitive to turnoff stars in an old stellar population and such that
fainter galaxies were older).  The bottom panel in Figure 3 indicates that
the $bz-yz$ CMR for A1185 is, in fact, confused and a blend of galaxies at
the faint end.  There is a clear correlation from the brightest galaxies to
$M_{5500} = -17$ (see dashed line fit in Figure 3), and the slope of this
relation matches the expectations from the SED models and the $vz-yz$
slope.  But, below this luminosity, the relationship dissolves and the
scatter increases such that a formal fit is a line of zero slope.  A
similar effect is noted by Poggianti \etal (2001) for giant versus dwarf
galaxies in Coma.  Also note, that due to this distortion in the $bz-yz$
CMR, our conclusions concerning the age of low luminosity galaxies in Coma
is invalid.  In fact, the poor resolution of the $bz-yz$ indicator to age,
motivated us to seek a more intensive procedure to determine age and
metallicity (see \S3.2).  which completely reverses our conclusions from
Odell, Schombert \& Rakos (2002) concerning the age of low luminosity
ellipticals.

\begin{figure}
\centering
\includegraphics[width=17cm]{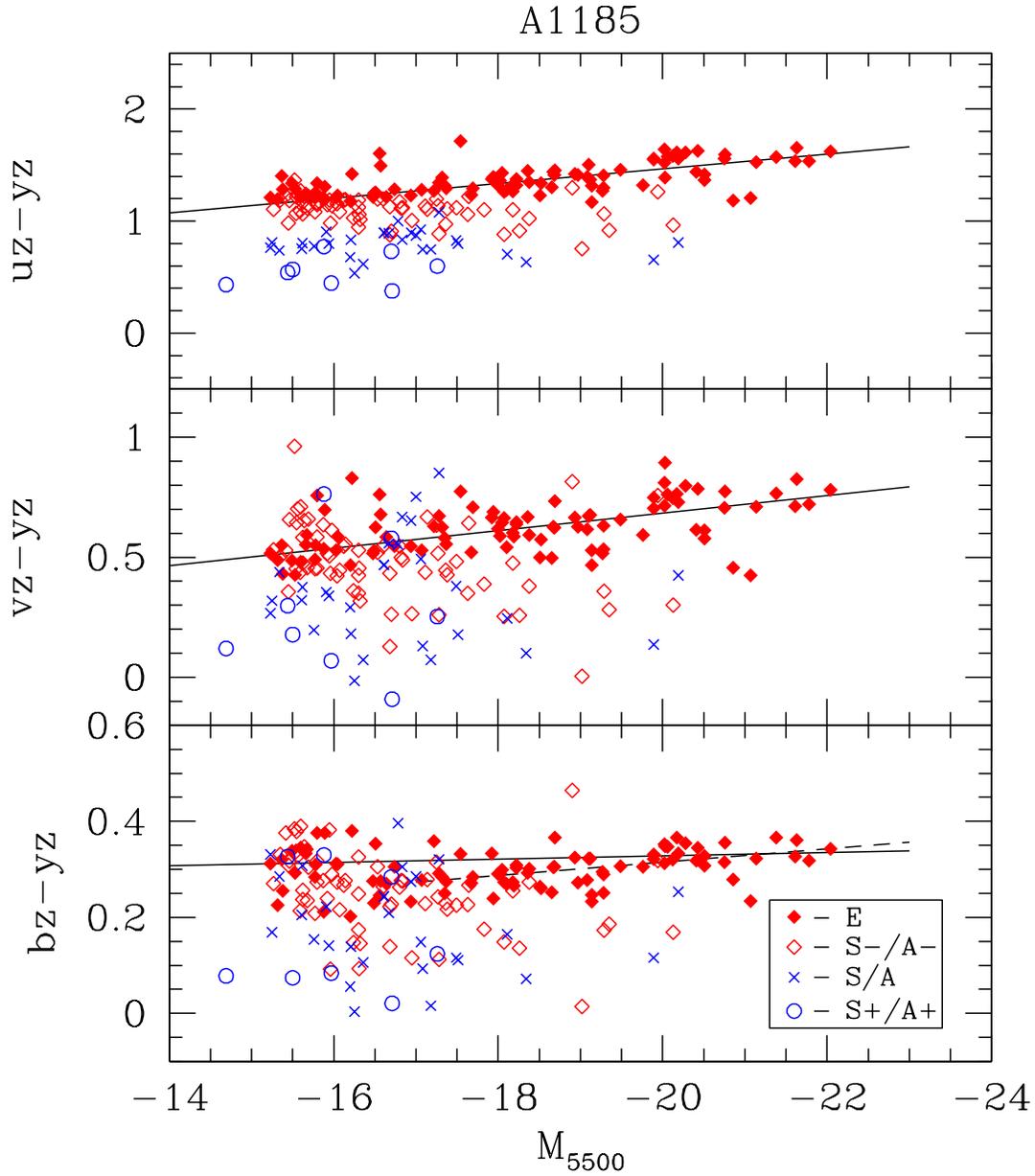}
\caption{The color-magnitude relation for A1185 in all three narrow band colors.
The solid line fits are from our Coma data (Odell, Schombert \&
Rakos 2002).  Both $uz-yz$ and $vz-yz$ display the linear correlation for red
galaxies that represents the run of metallicity with mass in ellipticals.  While the
Coma data indicated a flat CMR for the $bz-yz$ index, the fainter data shown
here for A1185 displays a normal CMR until $M_{5500} = -17$ then an upward trend
is visible which distorts the linear fit.  The dashed line indicates a linear
fit to the $bz-yz$ data for red galaxies brighter than $-17$.
}
\end{figure}

The data for A1185 also indicates there exists a spectrophotometric-density
relationship in A1185, equivalent to the density-morphology relation
(Dressler 1984, Smith \etal 1997).  Figure 4 displays the variation on
projected galaxy density as a function of cluster radius for the four
photometric classes (S and S+ were combined to represent all star-forming
galaxies).  Unsurprisingly, star-forming galaxies avoid the cluster core
and the red population (E and S- classes) dominate the inner regions.  Numerous
environmental effects have been postulated to account for rapid changes in the
star formation of disk galaxies as they encounter a cluster core such as
tidal forces, gas stripping, strangulation (see Poggianti 2004 for a review). 
However, even the S- class displays a statistically significant drop in
density at the cluster core, even though there are no overt signs of star
formation in these systems.  An environmental effect on galaxies with old
stellar populations would imply that the cluster environment has had a long
history on galaxy populations, and that even within the red population of a
cluster there are color differences influenced by environmental factors.

\begin{figure}
\centering
\includegraphics[width=12cm,angle=-90]{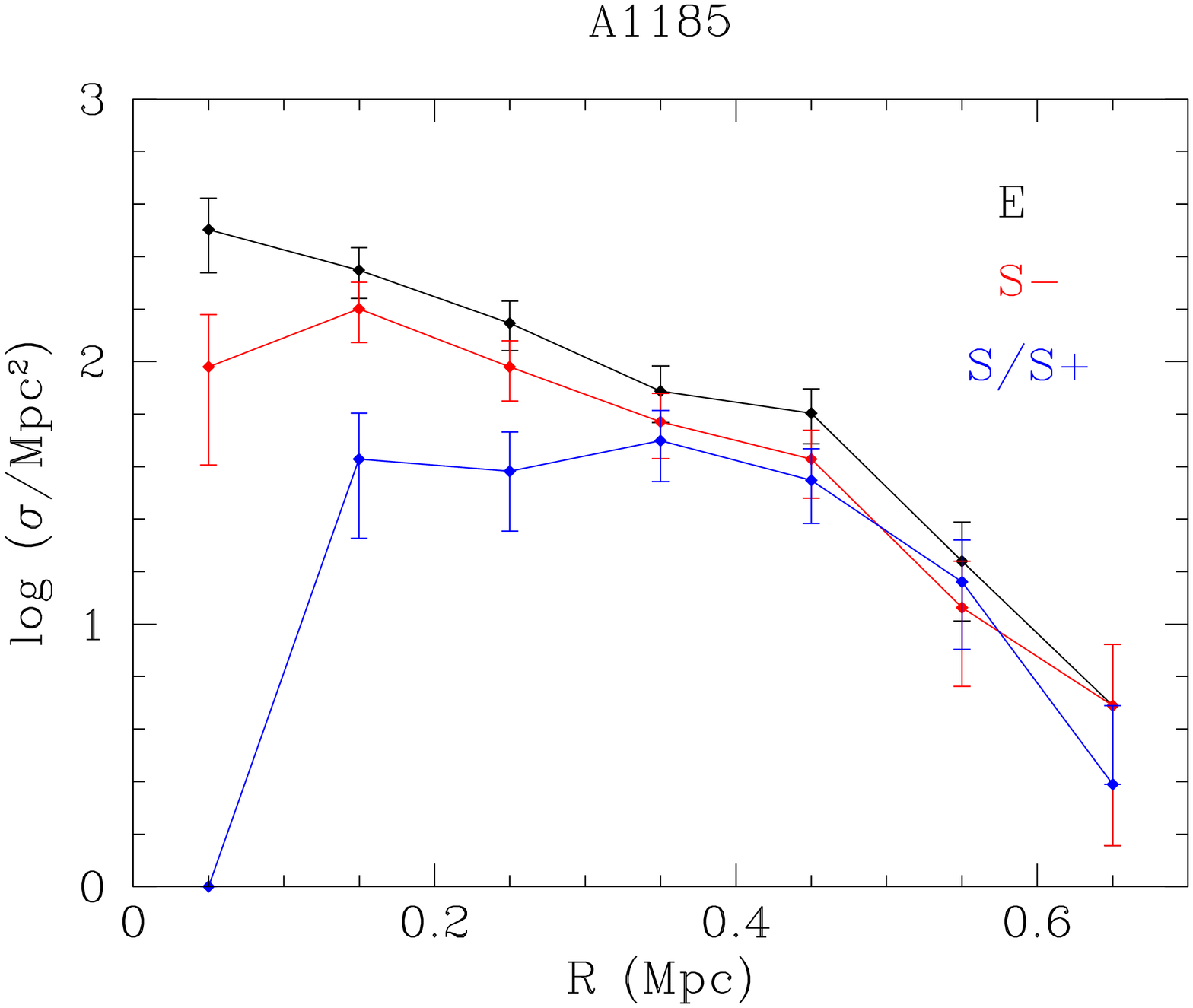}
\caption{The project radial galaxy density for each spectrophotometric class as
a function of distance from cluster center.  There is a noticeable drop in star
forming galaxies towards the cluster core, presumably an environmental effect on
the star formation history of galaxies.  However, even red S- galaxies display a
decrease into the cluster core.
}
\end{figure}

\subsection{Age and Metallicity Determination}

The goal for spectrophotometry studies of cluster galaxies is to reveal
some information pertaining to their star formation history.  Ultimately,
we wish to determine not only the formation epoch of the galaxies (the age
of their oldest stars), but also the style and behavior of their star
formation (for example, initial bursts versus later episodes) as it factors
into their evolution.  The optical colors of galaxies reflect information that
pertains to the photospheres of stars and, therefore, the evolution of a
galaxy's colors mirrors changes in those same stellar populations.  Thus, the
light from stellar populations in galaxies is primarily a two parameter
family based on the distribution of age and metallicity of the stars.

Techniques for estimating the age and metallicity of galaxies have improved
dramatically in the last 20 years.  For optical studies, the primary
technique of choice is age and metallicity determination through the use of
spectral features (Gonzalez 1993).  This has the immediate advantage of
linking galaxy measurements to stellar studies, where a solid foundation of
theoretical work on stellar atmospheres has resolved the behavior of
spectral lines with temperature and abundance.  However, the dominant
component that produces the observed colors in galaxies is not their
spectral features but rather the continuum emission.  Age and metallicity
vary the continuum emission through the impact of higher mass stars in
younger stellar populations (hot colors) and changes in the color of the
red giant branch (RGB) by line blanketing effects caused by changes in the
mean metallicity (Tinsley 1980).

Stellar population studies by colors is not without drawbacks.  For example,
the well known degeneracy problem for continuum colors derives from the
fact that changes in metallicity can imitate age effects, particularly in
observations without sufficient spectral resolution (i.e. from broadband
colors such as Johnson $UBV$).  Our use of narrow band continuum colors is
motivated by two factors; 1) narrow filters select smaller regions of a
galaxy's spectra, regions which contain a concentration of metal lines (the
$vz$ region) versus regions that are relatively free of line (continuum
areas, i.e. $bz$ and $yz$) and 2) there exist improved population models
(Schulz \etal 2002) which sample the region of age-metallicity parameter
space useful for the analysis of old, red galaxies and which are now
tied to the globular clusters system.  The Schulz \etal
models are a significant improvement over past work with respect to our
near-UV and blue colors since they 1) include the recent isochrones from
the Padova group which have a stronger blue HB contribution, 2) use more
recent stellar atmosphere spectra which, again, are refined to better
sample the blue and 3) are in better agreement with globular cluster $uvby$
colors than previous work.  Our studies on the use of the $uz,vz,bz,yz$
filters to determine age and metallicity have culminated in Rakos \&
Schombert (2005a) where we develop a principal component methodology for
determining these two parameters in composite systems, such as galaxies.
There are several limitations to the technique.  For example, the models
are restricted to [Fe/H] $>$ -1.7 and ages greater than 3 Gyrs.  However,
for cluster galaxy studies, we can isolate galaxies with ongoing star
formation, or have large fractions of young stars, by their extreme colors
and/or strong emission lines.  In addition, there is a strong expectation
that cluster galaxies consist of star populations with mean [Fe/H]
values greater than globular clusters, removing any problems of fitting the
low metallicity end of the models.

Our approach derives metallicity and age values using only the observed
narrow band colors and the knowledge of the behavior of the PC surface
(i.e., guided by the SED models).  We first use the three narrow band
colors from the SED models (Schulz \etal 2002) to generate a mesh of PC1
and PC2 values, where the PC equations (see Rakos \& Schombert 2005a)
contain age, metallicity and the three narrow band colors.  With the
expectation that one pair in the computed mesh represents the true value
for an unknown age and metallicity, we next attempt to identify the true
value of PC1-PC2 for an object with observed colors but unknown age and
metallicity by an iterative search.  The search begins by using an
estimated metallicity from the $vz-yz$ color and a crude, mean stellar age
from the $bz-yz$ color.  We then interpolate a PC1 and PC2 pair from the
assumed age and metallicity, use the PC equations to fill in the observed
color indices plus the assumed age and metallicity, and compare the
calculated pairs to those derived from the colors.  In practice, the search
for the correct age and metallicity is an iterative process as the PC plane
is smooth, but not linear.  Correct values are evaluated by calculating the
differences between the PC values determined from models and those
determined from the PC equations, in which the observed colors are the
input. The quality of the solution is measured by the root mean difference
(rms) between the PC values over two iteration loops; one for age and one
for metallicity.  At each step, PC values are interpolated from the Schulz
et al. (2002) models for the step values of age and metallicity, then
compared to the PC values determined from the PC equations, again using the
step values of age and metallicity plus the observed colors.

There are several cautionary notes to this technique: 1) the procedure can
only be used to determine mean ages and metallicities (i.e., objects which
have multiple epochs of star formation will produce age/[Fe/H] values that
are the luminosity weighted mean), 2) the technique only applies for
stellar populations older than 3 Gyrs (i.e., the degeneracy is still
unresolved for galaxies undergoing starburst or recent star formation), 3)
the age/[Fe/H] values are integrated values from the total luminosity of
the galaxy and are not surface brightness selected (e.g. core
spectroscopy), this means that some metallicity/age model must be used to
interpret internal stellar population distribution of age and metallicity
as it reflects into the integrated colors.  The tools for our new
age/metallicity algorithms can be found at
http://student.fizika.org/$\sim$vkolbas/pca.

\begin{figure}
\centering
\includegraphics[width=17cm]{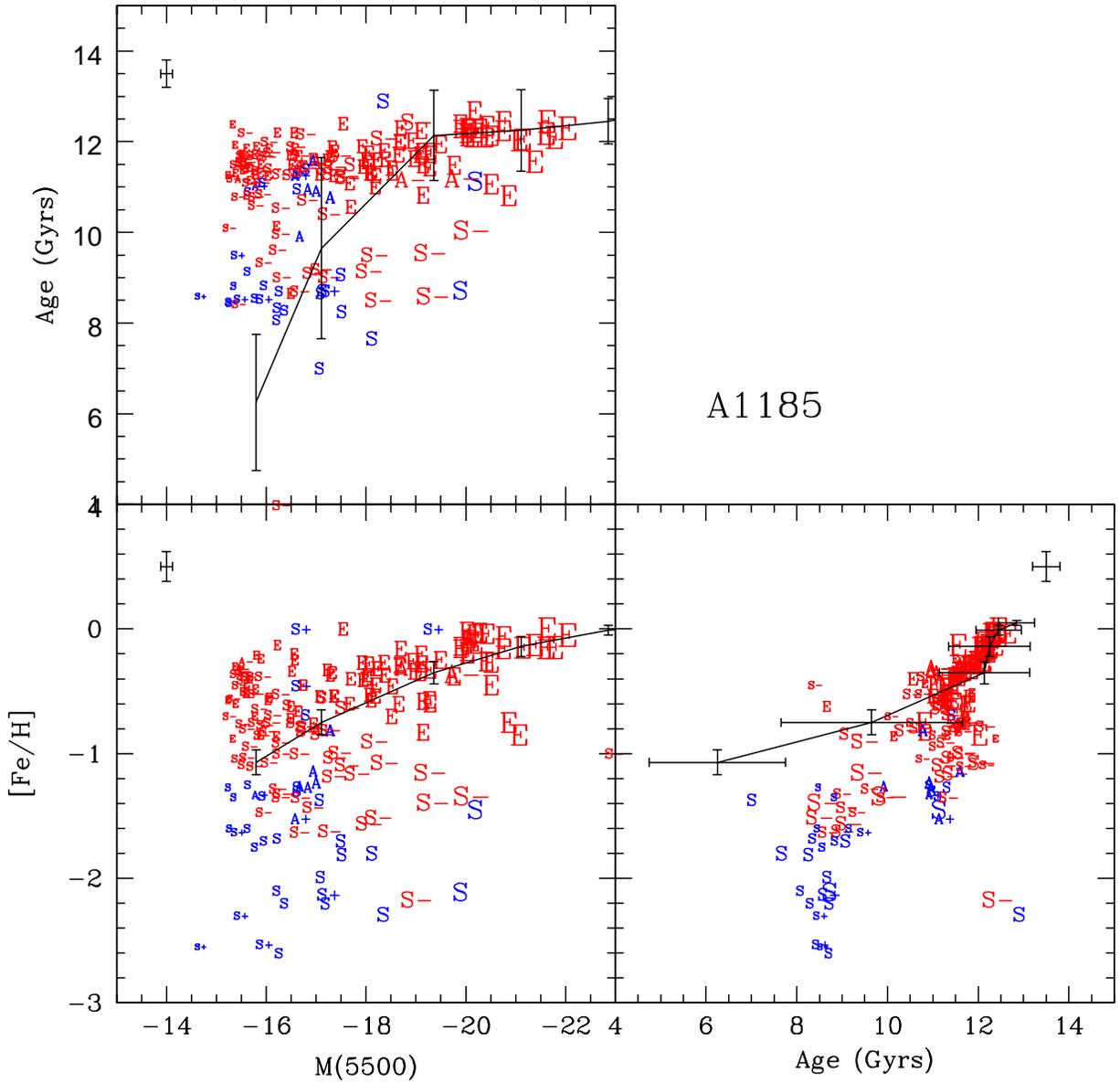}
\caption{The age, metallicity, mass (luminosity) diagram for A1185.  This
diagram compares the photometrically determined age and metallicity for all 192
members of the A1185 cluster.  Each datum is labeled by its spectrophotometric
classification.  Two distinct groups are visible in the age-luminosity frame,
one with a mean age of 12 Gyrs and a second group with a mean age of 9 Gyrs.
The mass-metallicity relation is visible for the brightest red galaxies (bottom
left panel), but degrades quickly at luminosities less than $-18$.  There is a
strong correlation between age and metallicity for the reddest galaxies (bottom
right panel) with the oldest galaxies have the highest integrated [Fe/H] values.
Also plotted as solid lines are the hierarchical galaxy formation models of De
Lucia \etal (2006) where the oldest galaxies were adjusted upward in age by 2.5
Gyrs to match the ages of the oldest galaxies in our sample.  While there is
qualitative agreement between these simulations and the data, there are large
numbers of systems that range in age and metallicity not predicted by the
models.
}
\end{figure}

The resulting age and [Fe/H] values for the A1185 galaxies, as calculated
using the technique described in Rakos \& Schombert (2005a), are listed at
our data website (http://abyss.uoregon.edu/$\sim$js/narrow) and displayed
in Figure 5.  The typical error in our technique is 0.2 dex in [Fe/H] and
0.5 Gyrs in age which, of course, degrades at lower luminosities.   What is
immediately obvious, in both the Coma and A1185 data, is that cluster
galaxies divide into two populations by age and [Fe/H].  The first
population is a classic red sequence of galaxies with ages greater than 10
Gyrs having a color-magnitude relation which is clear for high luminosity
systems, but degrades at lower luminosities.  Their morphological and
spectrophotometric classifications are early-type ellipticals and S0's with
no current or recent star formation.  The metallicities display strong
chemical evolution with values that range from [Fe/H]$=-1.5$ to slightly
greater than solar (see also Odell, Schombert \& Rakos 2002).  The second
group of cluster members is younger than the first group with mean ages
between 8 and 10 Gyrs (although for some of the star-forming members of
this group, this age will represent an upper limit).  Their metallicities
also reflect a weakly evolved history with mean [Fe/H] values less than
$-1$.  An advantage to our narrow band color scheme is that these [Fe/H]
values are tied to the SED models.  As there are significant changes in
$\alpha$/Fe ratios with galaxy mass (Sanchez-Blazquez \etal 2006), [Fe/H]
extracted from line values must correct for this effect.  Comparison to
super-rich metallicity models (Rakos \etal 2001) shows that changes in
Mg/Fe from 0 to 0.4 result in changes in $vz-yz$ color of less than 0.01
mags.

The old, red group displays a classic trend of luminosity and [Fe/H] for
galaxies brighter than $-$18, but the relation degrades at lower
luminosities.  Assuming that luminosity traces stellar mass (Cappellari
\etal 2006), then bright red galaxies follow a well-defined
mass-metallicity relation (Kelson \etal 2006, Nelan \etal 2005).
Comparison of all three narrow band colors also demonstrates that the CMR,
in our color system, is primarily a metallicity indicator.  The effects of
age are secondary, although in the same direction as metallicity effects
with galaxy luminosity.  The low scatter in the CMR for red galaxies is not
due to competing age and metallicity changes.

This older population also displays a clear correlation between age and
luminosity (stellar mass) in the sense that more massive galaxies are
older.  This trend indicates a mean age for high mass ellipticals to be
12.5 Gyrs, on average, dropping to 11.5 Gyrs for the faintest ellipticals in
the sample.  This effect is often referred to as the `downsizing' scenario of
galaxy formation (Smail \etal 1998, Bundy \etal 2005, Mateus \etal 2006)
and has been confirmed in line indices work (Collobert \etal 2006, Gallazzi
\etal 2006, Sanchez-Blazquez \etal 2006, Gallazzi \etal 2005, Thomas \etal
2005, Kuntschner \etal 2001), optical colors (Cool \etal 2006), near-IR
colors (James \etal 2006), UV colors (Kaviraj \etal 2006), $M/L$ estimates
(di Serego Alighieri, Lanzoni \& Jorgensen 2006), field galaxies (Li, Zhang
\& Han 2006, Mateus \etal 2006), cluster galaxies (Nelan \etal 2005,
Poggianti \etal 2004, Poggianti \etal 2001) and galaxies at higher
redshifts (Kelson \etal 2006, Mei \etal 2006, Schiavon \etal 2006, Barr
\etal 2005, Jorgensen \etal 2005).  The best demonstrations of the
downsizing effect are the CMR diagrams from $z=0.7$ clusters by De Lucia
\etal (2004).  These diagrams show a clear deficiency of low luminosity red
galaxies between $M_V=-19$ to $-17$.  Our studies of the Butcher-Oemler
effect in clusters, using the same narrow band filter set (Rakos \&
Schombert 2005b), finds that only 5 to 10\% of a galaxy's mass needs to be
involved in star formation to produce spiral-like colors as seen in the De
Lucia \etal diagrams.  Given that $z=0.7$ corresponds to 6 Gyrs ago
(Benchmark cosmology), then this would produce a current generation of low
luminosity cluster ellipticals with approximately 10\% of their stellar
mass in a 6 Gyr population plus the remaining 90\% of their stellar mass in
a 12 Gyr population.  Convolving these values to our SED models finds that
the luminosity weighted value for the age of such a galaxy would be 11
Gyrs, in agreement with the age estimates from Figure 5.

The second group of galaxies, with younger mean ages, is represented in
A1185 by two types of galaxies, a disk/irregular galaxy mixture on the low
mass side combined with a few S- class objects on the high mass side.  In
Coma, the younger population is primarily red, dwarf elliptical type
systems (see Odell, Schombert \& Rakos 2002 for more discussion).  The low
mass red galaxies in younger group are primarily non-nucleated dwarf
ellipticals, which we have found be have many of the characteristics of the
blue globular clusters in ellipticals (i.e. slightly younger and metal-rich
compared to Galactic globulars, Rakos \& Schombert 2004).  This supports
our claim in Rakos \& Schombert (2004) that low mass ellipticals divide
into two populations, nucleated dwarfs with ages and metallicities that
form a continuum with bright ellipticals and a younger non-nucleated dwarf
elliptical population with slightly younger ages and higher metallicities.

Our interpretation of blue galaxies in this second group, given their blue
colors indicating ongoing star formation, is a population with a majority
of their stellar mass in an old stellar population and a smaller mass,
young (less than 1 Gyr) stellar population which combine to produce an
integrated age of 8 to 9 Gyrs.  Again, invoking estimates using SED models,
we find that about 10\% of the mass of these blue galaxies can be in a one
Gyr population and the remaining in a 12 Gyr to produce the observed colors
and ages.  It is interesting to note that the high mass systems in the
second group appear to have the oldest ages, reddest colors and perhaps
represent a transition population from the blue to red cluster galaxies
(possible future S0's).  This population maintains many of the
characteristics for the `blue-side' population found in SDSS CMR studies
(Cool \etal 2006, Kauffmann \etal 2003), galaxies with primarily red
stellar populations and a small percentage of stars with ages less than one
Gyr resulting in a dichotomy in color space by galaxy mass.

Other points to note from Figure 5, the correlation between luminosity and
metallicity for the red population is weaker than the correlation between
mean age and metallicity.  While a mass (luminosity)-metallicity
correlation is evident in the data, in fact, for values of [Fe/H] above
$-0.5$, the metallicity of a red galaxy so well correlated with age that
either parameter predicts the other.  The correlation between age and
metallicity, in a positive sense, is in contradiction with other studies
such as Thomas \etal (2005) and Sanchez-Blazquez \etal (2006) which find
anti-correlations between age and metallicity (yet, positive correlations
between galaxy mass and age or metallicity).  Correlated errors between the
determination of age and metallicity could lead to positive correlation
seen in Figure 5 since redder colors can signify older age or higher
metallicities.  However, examination of the least square fits to the
parameters in PC space (see Figure 4 of Rakos \& Schombert 2005) finds that
$\Delta(Fe/H)$/$\Delta(age)$ has a slope of $-$0.03.  Whereas, the slope of
the age/metallicity relation in Figure 5 is $+$0.4.  Although the range in
age is low, it is unlikely that the correlation found from narrow band
colors is due to coupling of the two parameters.  Thus, Figure 5
demonstrates that the range of age and metallicity vary among galaxies,
particularly at low masses.  But for the oldest, high mass ellipticals, age
and metallicity have a strong linear behavior such that [Fe/H] = 0.36
$t_{form}$ - 4.55 in Gyrs.  The existence of this tight relationship for
the most massive galaxies is probably the strongest signature of a common
origin and evolutionary path for these systems.  Although the expectation
from simple galactic wind models (Arimoto \& Yoshii 1987, Matteucci 1994)
are that the oldest galaxies should have the lowest metallicities (older
meaning shortest duration of star formation and less processing of the
ISM), the reverse as seen in Figure 5.  This will be discussed further in
the conclusions section.

A comparison between our spectrophotometric classifications and galaxy age
is found in Figure 6.  The two age groups are clearly visible in this
normalized histogram with mean ages at 12 and 8.5 Gyrs.  Also noticeable in
this histogram is that even within the two groups there are differences in
ages by spectrophotometric type.  The median age drops from 12 Gyrs for E
class galaxies to 11.5 Gyrs for S- class galaxies to 11 Gyrs for
star-forming S/S+ class galaxies.  For the later type galaxies, this
probably reflects a mean age value based on a mixture of an old ($>$ 11
Gyrs), bulge stellar population plus a younger disk population.   However,
for the non-starforming S- class galaxies, their colors are typically more
uniform and we would not propose that the younger age is due to a mixing of
disk and bulge populations.  Yet, their ages from integrated light
measurements are statistically younger than the E class galaxies.  The
second, younger group is composed solely of S-, S and S+ class galaxies.
Where the youngest galaxies are, not surprisingly, the star-forming S and
S+ galaxies.  However, the distribution in age for S- galaxies is
distinctly bimodal with the youngest S- galaxies being older as the S/S+
galaxies (9.5 to 8.5 Gyrs), but certainly separate from the peak at 11.5
Gyrs.

The meaning of the trend with age and stellar mass is more difficult to
extract from our data because the integrated age of the underlying
population can be reached by many avenues.  For example, if all the stars
in a galaxy are formed simultaneously, then our age measurement represents
the time to that epoch.  This is also true if galaxies form their initial
stars over a range of time, but all starting and stopping at the same
epoch.  Then our age measurement will represent the luminosity weighted
mean age of the stars, presumably near the time to the median of the
initial star formation epoch.  This enters into a debate of the formation time
of the stellar population versus the assembly time of those populations into
their parent galaxy (De Lucia \etal 2006).  Under a monolithic collapse
scenario, those times are nearly equal.  But, recently, a variety of
hierarchical models have been considered where stars forming in smaller units,
then merger into their final systems.  One such set of models is presented by De
Lucia \etal (2006) using the results of a high resolution cosmic structure
growth simulation (the Millennium Simulation, Springel \etal 2005).  Their
resulting tracks of age, metallicity and galaxy mass (converted to
$M_{5500}$ using an $M/L$ of 4) are shown in Figure 5.  The error bars on
the models display the range from their simulations, not inaccuracies to
their method.  While there is qualitative agreement with the simulations in
that age increases with galaxy mass and metallicity increases with mass,
the simulations appear incomplete at the low mass end and even the broad
range of age in their low mass galaxy simulations do not capture the
distribution of ages and metallicities on our sample.  In particular, this
flavor of models fails to find old, yet metal poor galaxies that represent a
majority of the galaxies fainter than $-$17.

As an aside, an alternative interpretation of the luminosity-weighted ages
arises if the duration of the epoch of initial star formation varies with
galaxy mass.  For example, the red population trend in Figure 5 may suggest
that the initial phase of star formation is shorter for high mass galaxies
than low mass ones.  An extended phase of initial star formation results in
a mean age that is younger than the age calculated from a short, intense
burst.  Thus, the red population in A1185 and Coma either 1) have low
mass galaxies to be younger with respect to their initial star formation
epoch or 2) have similar formation epochs, but low mass galaxies experience
a longer duration of initial star formation as compared to higher mass systems.
This later interpretation is supported by the $\alpha$/Fe ratios in ellipticals
(Denicolo \etal 2005, Sanchez-Blazquez \etal 2006), which increase with
increasing galaxy mass to indicate a shorter duration of star formation.

\begin{figure}
\centering
\includegraphics[width=12cm,angle=-90]{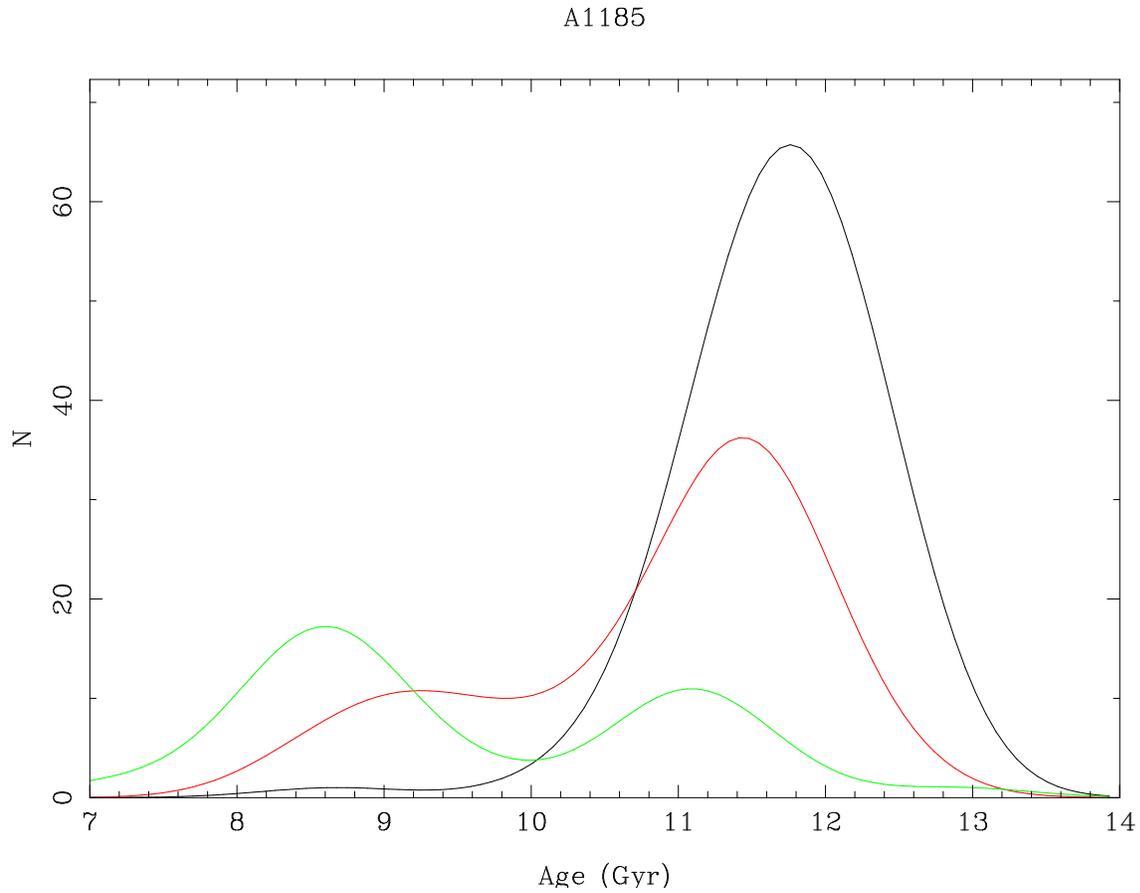}
\caption{Normalized histograms of age for the three spectrophotometric
classification groups: E (black line), S- (red line) and S/S+ (green line).
All three histograms are bimodal with secondary groups between 8 and 9 Gyrs
in age.  For the oldest galaxies in each class, there is a trend for
younger mean age with decreasing spectrophotometric type, presumably the
influence of the last epoch of disk star formation to the integrated
colors.
}
\end{figure}

Evidence for recent star formation in elliptical (Barger \etal 1996) would
complicate the interpretation of our ages calculated from integrated
colors.  These claims are based on spectroscopic signatures in the Balmer
series (i.e. H$\delta$).  In an old population, very small numbers of hot,
young stars are easy to detect in spectra.  Estimating the mass of a galaxy
involved in a recent burst has not been conclusive.  Our estimates from SED
models is that a recent burst, which involves less than 5\% of the galaxy
mass, will effect on our narrow band colors at less than the level of the
internal errors (Rakos \& Schombert 2005a).  Thus, there is no conflict
between observation of recent, weak bursts of star formation and our age
estimates which will apply to a majority of the underlying stellar population.

\subsection{Cluster Age Gradients}

Our Coma sample (Odell, Schombert \& Rakos 2002) lacked sufficient sky
coverage to confidently address correlations related to distance from
cluster center.  A larger field of view for the A1185 observations allows
us to address those issues in this study.  Figures 7 and 8 display just
that information for A1185, age and metallicity as a function of distance
from the geometric center of the cluster.  There are weak, but
statistically significant, trends for both parameters in the sense of
decreasing metallicity and decreasing age with increasing distance from the
cluster core.  These gradients are identical in direction and similar in
magnitude to the line index gradients found in the NOAO Fundamental Plane
Survey (Smith \etal 2006).  They also found a decrease in age by 1 Gyrs (15\%)
and almost no metallicity gradient (2\%).  Their interpretation is later
accretion produces younger stellar populations in a cluster halo, but early
environmental effects, such as quenching, could equally well explain the
observed gradients (Poggianti \etal 2004, Bundy \etal 2005).

As we have already seen in previous section, there is a
correlation between galaxy mass (luminosity) and age/metallicity.  So the
trend with distance from the cluster core may simple reflect mass
segregation.  However, an examination of the galaxy luminosities as a
function of cluster radius finds no correlation between galaxy luminosity
and radial distance.  The change in mean stellar age is approximately 1 Gyr
from the cluster core to the outer regions.  This would represent a change
in 2 to 3 mags in average galaxy luminosity as given by Figure 7, which is
simply not seen in the data.  Thus, the trends of age and metallicity are
not due to changes in galaxy mass with cluster position.

\begin{figure}
\centering
\includegraphics[width=17cm]{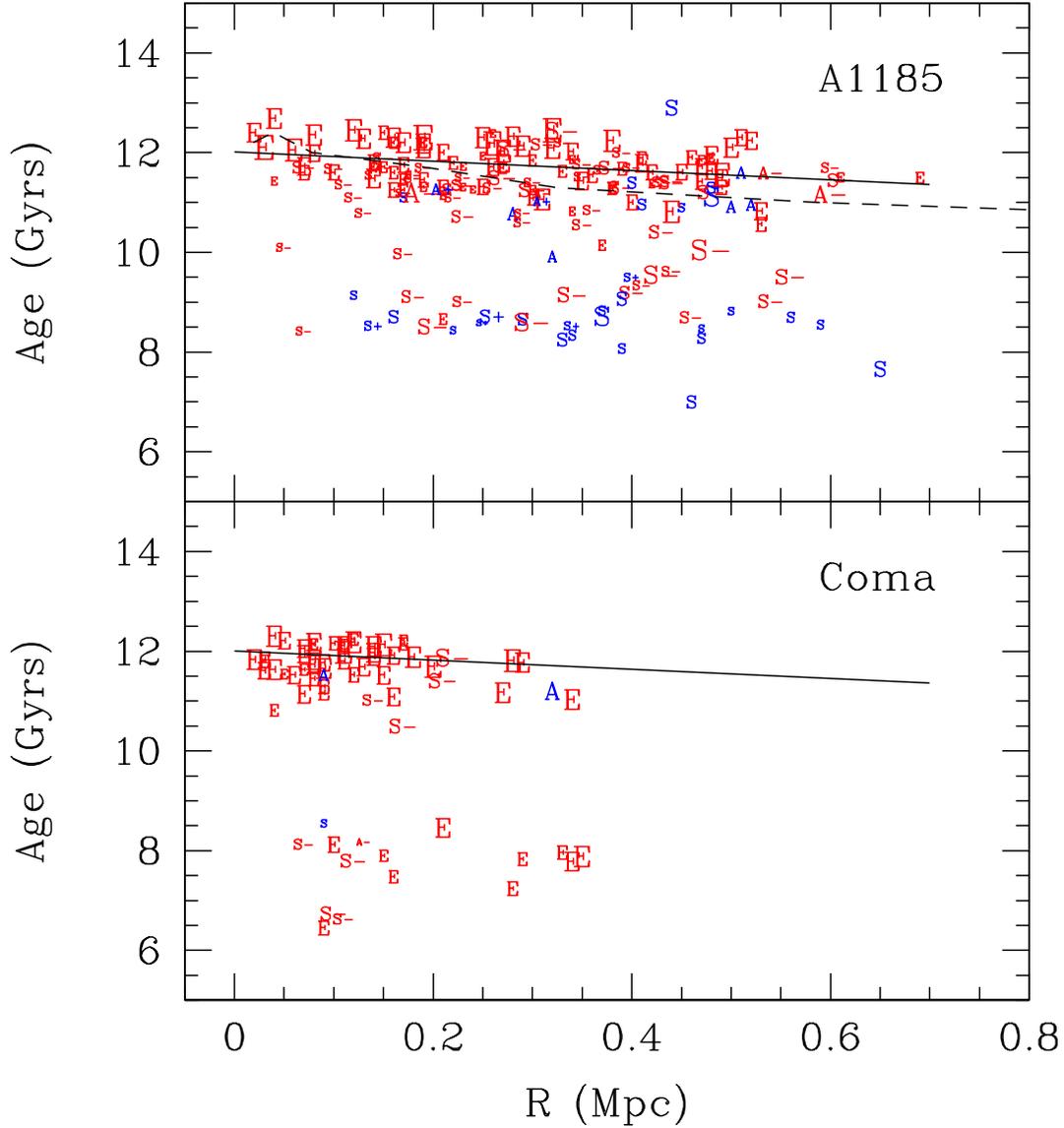}
\caption{The radial change in galaxy age with cluster distance for A1185 and
Coma.  The solid line is a best fit to the red population of A1185.  There is a
clear trend of decreasing age with increasing cluster distance.  
The dashed line
is from the hierarchical galaxy formation models of De
Lucia \etal (2006), which also predict an age gradient in rich clusters of
similar magnitude as seen in our data.
}
\end{figure}

Assuming that the trend of age with distance from the cluster core is real,
and not reflecting some secondary correlation with a galaxy's properties,
we are surprised to find that metallicity-radius relation is so poor
compare to the age-radius trend.  For the red population in A1185, the
relationship between age and [Fe/H] is very tight (see \S3.2).  An age
gradient of 1 Gyr should be matched by a corresponding metallicity gradient
of 0.5 dex, which is not seen in Figure 8.  This would agree with line strength
studies (Bernardi \etal 2006) which also found little change in Fe
indicators across various galaxy density regions.  While low metallicity
systems appear to avoid the cluster core, high metallicity galaxies
([Fe/H]$>-0.5$) are found at all cluster radii.  The implication here is
that cluster environmental effects, which alter the red populations star
formation history, affect the star formation rate (i.e. age) more sharply
than the chemical evolutionary history of a galaxy, i.e. the mean
metallicity of a red galaxy is set long before the era of star formation is
complete.

\begin{figure}
\centering
\includegraphics[width=17cm]{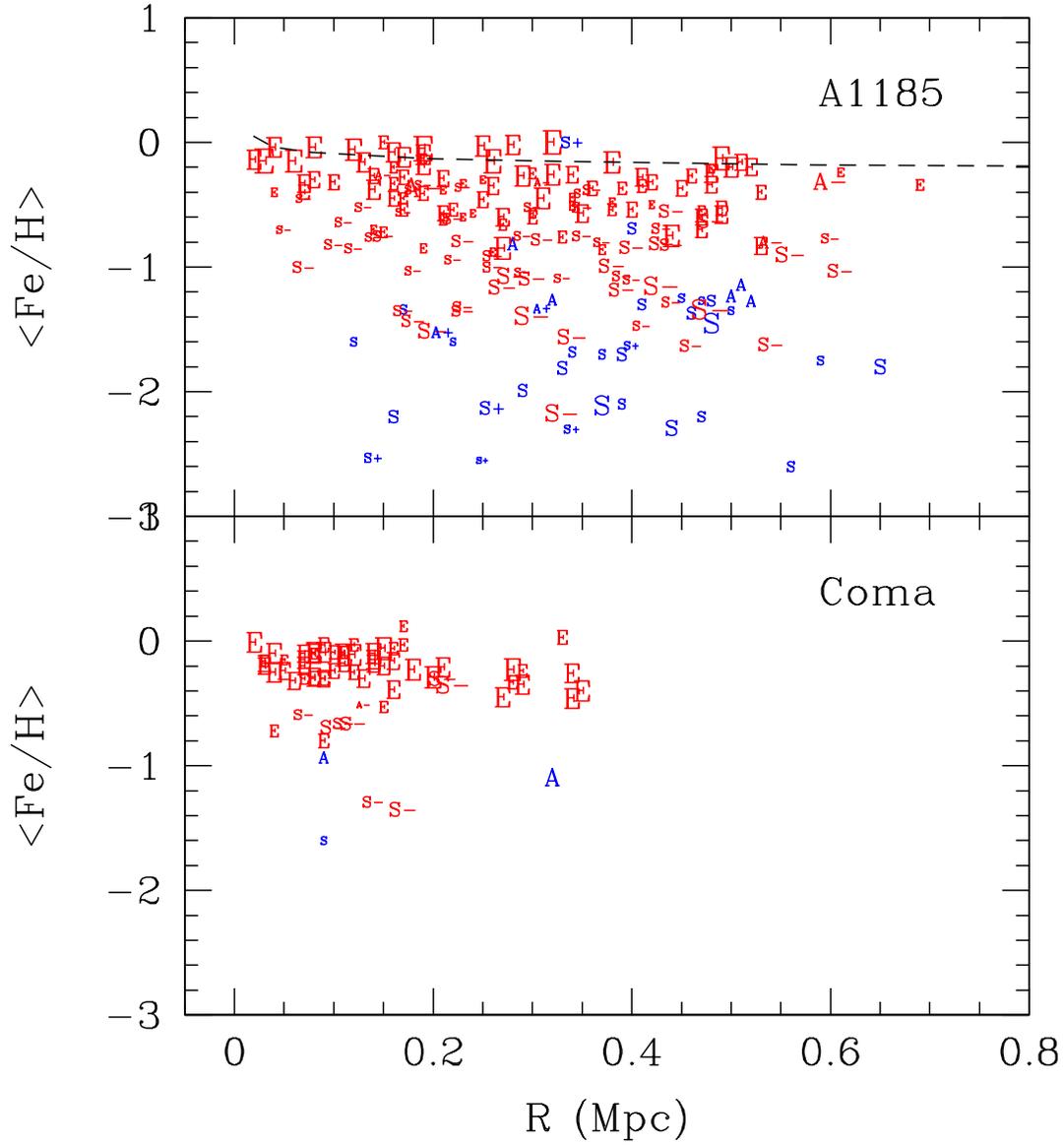}
\caption{The radial change of metallicity ([Fe/H]) with cluster distance.  As
with Figure 7, there is a trend of decreasing galaxy metallicity with cluster
distance.
The dashed line
is from the hierarchical galaxy formation models of De
Lucia \etal (2006), which predict a small metallicity gradient in rich clusters.
}
\end{figure}

Various hierarchical models of galaxy formation predict differences between
the ages of galaxies formed in high density versus low density regions
(Baugh, Cole \& Frenk 1996).  Present-day clusters form in high density
peaks of the primordial density field.  This leads to early collapse times
for galaxies forming in these regions and, therefore, an older stellar
population compared to field galaxies.  In addition, there exist
simulations that display age and metallicity gradients within the cluster
environment based on the hierarchical recipes (De Lucia \etal 2006, shown
as dashed lines in Figures 7 and 8).  There is some evidence for this
effect in spectroscopic studies (Bernardi \etal 2006, Collobert \etal
2006); however, the data presented herein does not sample into the field
population as would be required to test this observation.  

The age gradients seen in Figure 7 follow the change in current galaxy
projected density from core to cluster halo, but these outer density values
are still a factor of three greater than the field, and there is no
guarantee that these galaxies are occupying the same local density today as
at their formation.  Gradients found in simulations (De Lucia \etal 2006)
are primarily driven by mass segregation such that the distance to the
center of a cluster is correlated loosely with time and, thus, is reflected
in the mass correlated properties of the individual galaxies.

\section{CONCLUSIONS}

The processes, both internal and external, that influence the star
formation history of cluster galaxies are so complex and interconnected,
that even the determination of relatively accurate values for mean age and
metallicity will not uniquely map into a single galaxy formation scenario
due to the large number of possible paths a stellar population may take to
end up with the integrated values that we measure.  The best we can do is
isolate the pure observables in this study and list the most likely
candidates for the processes that a galaxy might take to acquire its
properties. This will, in the end, be a list of the most likely star
formation scenarios combined with the known environmental influences on
cluster galaxies.

With respect to the red population in rich clusters, our work herein has
found clear trends in galaxy age and metallicity with respect to stellar
mass and environment.  The most massive galaxies in a cluster are the
oldest with luminosity weighted ages of 12 to 13 Gyrs.  These galaxies also
have the highest metallicities, near solar.  With decreasing stellar mass
(luminosity), the mean age decreases to approximately 11 Gyrs.  Likewise,
the mean metallicity decreases to [Fe/H]=$-0.5$.  Age and metallicity are
strongly correlated, such that older mean age correlates to higher mean
metallicities.  In addition to trends with galaxy mass, the mean age of the
red population also changes with environment with the oldest galaxies found
in the dense inner core and the youngest galaxies at the lower density
cluster halos, although there is little change in global metallicity with
cluster location.

As stated in \S3.2, it is impossible to determine strictly from integrated
values whether the variation in mean age with galaxy mass is due to varying
epochs of initial star formation or varying durations.  However, an
additional stellar clock is available to resolve this dilemma, the ratio of
$\alpha$ elements to Fe.  The ratio of $\alpha$ elements to Fe is a
function of the number of Type II versus Type Ia supernova in a galaxy's
past.  Type Ia SN produce extra amounts of Fe and decrease the ratio.
However, Type Ia SN require at least a Gyr of time to build-up within a
galaxy and, therefore, high fractions of $\alpha$/Fe indicate shorter
duration times for star formation.  SDSS (Sanchez-Blazquez \etal 2006) and
deep line indices studies (Nelan \etal 2005) have already demonstrated that
$\alpha$/Fe increases with galaxy mass (although see Matteucci \& Chiappini
2005 for a dissenting view on the $\alpha$ abundance problem).  The
relationship of age with galaxy mass seen herein is consistent with a
single epoch of galaxy formation and increasing initial durations of star
formation with decreasing stellar mass.  This appears to be a weaker
version of the so-called `downsizing' effect discovered in other age
studies (Gallazzi \etal 2006) since, by comparison to SED models, if as
much as 10\% the stellar mass of a low luminosity elliptical is involved in
star formation at $z=0.7$ (De Lucia \etal 2004) then this would produce
luminosity-weighted mean ages of around 11 Gyrs at the present epoch.

The homogeneous properties of the red population, for example, the tight CMR
found in all rich clusters, has always argued strongly for a monolithic
collapse scenario (Partridge \& Peebles 1967, Larson 1975).  Yet, the
decrease in age with galaxy density (radial changes) found in this study is
precisely the effect predicted by most hierarchical formation models (Yee
\etal 2005, Conselice 2006) and, more importantly, an increase in
metallicity with galaxy mass in conjunction with a decreasing duration of
initial star formation with galaxy mass is in direct contradiction of the
classic galactic wind model of galaxy formation.  Under a normal wind model
(see Matteucci 2004), the efficiency of star formation is a decreasing
function of galactic mass.  Thus, the timescale of star formation is based
on the dynamical free fall time.  Larger mass systems have longer dynamical
timescales and, therefore, smaller efficiencies and longer durations of
initial star formation before the onset of a galactic wind halts the
process.  This allows more time for chemical evolution to increase the
abundance of metals and resulting in higher global metallicities for higher
mass galaxies (i.e., the color-magnitude relation).  However, the
$\alpha$/Fe ratios and the age-mass relation (Figure 5) argue exactly the
opposite of the classic wind model, that star formation was shorter and
more efficient with increasing galaxy mass.  In order to maintain a
homogeneous collapse scenario, one is required to propose a process where
ellipticals form by merging protoclouds and the merger process introduces
higher densities/velocities resulting in more efficient, and more
abundant, star formation (i.e., the inverse wind model, Matteucci 1994).
This would be a hybrid galaxy formation scenario, mergers early on then
reducing by monolithic collapse to a final state.

Gradients exist in A1185, a dynamically young cluster.  But the gradients
are more discernible in age, than in metallicity.  Environmental effects
are usually thought to primarily influence the star-forming population in
clusters, we note that even the red population of S- class galaxies display
a gradient with respect to cluster density and effect that most have
applied at the earliest formation times.  These two observations would
indicate that gradients are suggestive of a `quenching' behavior for
cluster galaxies (Poggianti \etal 1999, Bundy \etal 2005) that has been in
place even as the cluster was formed at redshifts greater than 5.  Thus,
the age and metallicity gradients found in A1185 appear to be the effect of
environmental processes imprinted at early epochs onto the red population.
This type of behavior is visible in the latest hierarchical models produced
by collisionless cosmological simulations (De Lucia \etal 2006).  These
models give a qualitative fit to our observations (see model tracks in
Figure 5) if the oldest galaxies in the models have their ages increased by
2.5 Gyrs.  Key to those models is that the star formation epoch matches
formation redshifts of 5, but that galaxies assemble by mergers at
redshifts below 2.  Our data cannot address the assembly timescale for the
red population, which is an issue to be investigated by space imaging.

The second population in A1185 and Coma (with mean ages around 9 Gyrs, see
Figure 5) is associated with the blue population in clusters and, for higher
redshifts, the Butcher-Oemler effect.  While most of this population has a mean
age based on the luminosity-weighted sum of a small fraction of young stars
plus a bulk of older (12 Gyrs) stars, a significant fraction of this group
have non-star-forming colors (class S-) and probably represent transition
galaxies between the Butcher-Oemler population at high redshifts and their
descendants at the current epoch.

This blue population is also a candidate for the source of the metal-rich
globulars found in many bright ellipticals.  In Rakos \& Schombert (2004), we
re-analyzed the narrow band colors of M87 globulars from Jordan \etal (2002)
and found that the bimodal aspect of the M87 globulars was due to a old,
metal-poor population (blue globulars) plus a younger, metal-rich population
(red globulars), in agreement with many other studies of extragalactic globulars
(see Brodie \& Strader for a review).  We estimated the age of the red
population of globulars to be between 3 and 4 Gyrs younger than the blue
globulars.  Interestingly, this is similar to the difference in age between the
old cluster galaxies and the blue population in A1185 and Coma.  It suggests
that, under a merger scenario for globular clusters in ellipticals (Ashman \&
Zepf 1992), the younger galaxies in clusters can be a source for the younger
globulars in ellipticals.

\acknowledgements

Financial support from Austrian Fonds zur Foerderung der Wissenschaftlichen
Forschung and NSF grant AST-0307508 is gratefully acknowledged.  This
research has made use of the NASA/IPAC Extragalactic Database (NED), which
is operated by the Jet Propulsion Laboratory, California Institute of
Technology, under contract with the National Aeronautics and Space
Administration.  We also wish to thank the faculty and staff of the
University of Arizona's Steward Observatory for the time allocated to us
using 90Prime on the Bok Telescope.  This material is based upon work
supported by the National Science Foundation under Grant No. 0307508.

\end{document}